\documentclass{article}
\usepackage[utf8]{inputenc}
\usepackage{authblk}
\usepackage{amsmath}
\usepackage{comment}
\usepackage{float}
\usepackage{xcolor}

\usepackage[numbers,sort&compress]{natbib}

\DeclareMathOperator{\sech}{sech}
\usepackage[a4paper, total={7in, 10in}]{geometry}

\usepackage{hyperref}
\hypersetup{
    colorlinks=true,
    linkcolor=blue,
    filecolor=magenta,      
    urlcolor=cyan,
    citecolor=blue,
}

\newcommand{\myfigure}

\DeclareMathOperator{\sn}{sn}
\DeclareMathOperator{\cn}{cn}
\DeclareMathOperator{\dn}{dn}
\title{Nonlinear Dynamics in Periodically Distributed Fibre Amplification Systems}

\author[1,2]{Sunday O. Atuba}

\affil[1]{  \textit{ \small  School of Engineering Mathematics and Technology, University of Bristol, Bristol United Kingdom}}

\affil[2]{  \textit{ \small  School of Engineering, University of the West of England, Bristol United Kingdom}}

\affil{\small uu26802@bristol.ac.uk; \quad alternative email: sunday.atuba@uwe.ac.uk}

\author[3,4]{Boris A. Malomed}
\affil[3]{  \textit{\small  Department of Physical Electronics, School of Electrical Engineering, Faculty of Engineering, Tel Aviv University, Tel Aviv 69978, Israel}}
\affil[4]{  \textit{\small  Instituto de Alta Investigaci\'{o}n, Universidad de Tarapac\'{a}, Casilla 7D, Arica Chile}}

\usepackage{natbib}
\usepackage{graphicx}

\begin{document}
\date{}
\maketitle

{\Large \textbf{Abstract}}

We investigate effects of longitudinally varying nonlinearity on periodically distributed amplification systems in
optical fibres. Emphasis is made on configurations in which the nonlinear coefficient exponentially grows or decays
along the propagation distance, while the group-velocity-dispersion (GVD) coefficient is held constant. Such systems
are of practical relevance to the design of highly efficient nonlinear devices, including broadband light sources and
advanced amplification schemes, where precise management of the interplay between the GVD, nonlinearity, and
gain is essential. Adopting the known generic integrability condition, we derive exact analytical expressions
for the evolution of the optical field in the presence of the tailored nonlinearity profiles and periodically distributed
gain. The resulting periodic wavetrain solutions and their long-wave limit in the form of bright and dark solitons
provide insight into the controlled enhancement or suppression of nonlinear effects under realistic amplification
conditions. The robustness of these solutions is examined by introducing finite perturbations and directly evaluating their effects on the corresponding intensity profiles.
The structural stability of solutions against deviations from the integrability
condition is demonstrated too. Outcomes of this study contribute to a deeper understanding of the nonlinear pulse
dynamics in engineered gain-managed fibres and offer guidance for the development of novel optical systems, where
the spatial distribution of the nonlinearity is exploited to optimise the performance.


\section{Introduction}

The field of optical fibre communications has been profoundly influenced by the introduction and utilisation of optical solitons, i.e., self-sustained wave packets that travel without distortion, owing to the equilibrium between
the group-velocity dispersion (GVD) and Kerr nonlinearity of the fibres. Since the pioneering work of Hasegawa and Tappert (1973) \citep{hasegawa1973transmission}, which had predicted the existence of solitons in optical fibres, they have served as a fundamental example of how the nonlinearity, rather than
being a source of signal degradation, can serve to enable distortion-free signal transmission over long distances. The Kerr effect, which fundamentally underpins nonlinear phenomena in optical communications, explicitly describes how intense optical fields alter the effective refractive index \citep{dziedzic1981optical}. 
This effect manifests itself not only in crystalline and glassy media, but also in gases. It occurs in two primary forms: the DC, alias electro-optic, Kerr effect, and the AC, alias optical, Kerr effect. The former one is caused by the application of an external electric field to the medium, inducing a polarisation-dependent phase shift, proportional to the squared field’s strength, whereas the latter effect arises intrinsically, with the nonlinearity stemming from the instantaneous response of the material to
the propagating light wave, which induces a nonlinear polarisation of the material, thus effectively modifying the local refractive index \citep{dziedzic1981optical}.

Temporal optical solitons emerge as stable solutions to the nonlinear Schrödinger equation (NLSE), which governs the propagation of light in single-mode fibres. NLSE includes the group-velocity-dispersion (GVD) and Kerr terms,
the latter originating from the dependence of the effective refractive index on the optical intensity. The Kerr effect gives rise to the self-phase modulation, i.e., a phase shift proportional to the local optical power, whereas GVD causes broadening of temporal pulses. As predicted in the above-mentioned pioneering work \citep{hasegawa1973transmission}, the soliton appears when these two effects mutually lock in a stable balance \citep{agrawal2000nonlinear}. Mollenauer, Stolen, and Gordon (1980) had reported the first experimental observation of fundamental and higher-order solitons in fibres (the higher-order ones, alias breathers, are bound states of fundamental solitons), directly confirming that, as mentioned above, the nonlinearity can be used beneficially, rather than viewed as a source of impairment \citep{mollenauer1980experimental}. This discovery marked a milestone in the advancement of high-capacity long-haul optical communications. Unlike conventional non-return-to-zero signals, which degrade under the action of nonlinear distortions, such as the self- and cross-phase modulation (SPM and XPM), soliton-based systems harness the Kerr nonlinearity to preserve the
pulses’ integrity.

In practical systems, there are additional factors which disturb the soliton stability, such as fibre losses, amplifier noise, and higher-order effects, including the third-order dispersion and stimulated Raman scattering. In addition to
that, the amplified spontaneous-emission noise gives rise to the Gordon-Haus effect \citep{gordon1986random}, which induces a random walk of the solitons in the temporal direction, thus compromising the transmission performance. In wavelength-division multiplexed systems, cross-phase modulation and four-wave mixing between soliton channels create the inter-channel cross-talk that disrupts the multi-channel soliton propagation \citep{agrawal2000nonlinear, wai1990radiations}.To alleviate these impairments, the dispersion management (DM) technique
was introduced, in the framework of which alternating fibre segments with opposite GVD signs (anomalous and normal) sustain the robust propagation of soliton-like modes, while limiting the build-up of the nonlinear phase shift \citep{smith1996enhanced, malomed2006soliton, turitsyn2012dispersion}. The concept of DM solitons has helped to extend the capacity of the soliton transmission into the multi-terabit range by maintaining the balance of the accumulated GVD. Additional techniques, such as the optical phase conjugation and digital backpropagation, have been proposed to compensate for excessive nonlinear distortions and maintain the soliton stability in coherent fibre-optic systems \citep{ip2008compensation}.

{Recent studies have also demonstrated substantial progress in the construction of exact nonlinear-wave families and the analysis of their stability in dispersive nonlinear systems, including nonlinear Schrödinger- and complex Ginzburg–Landau-type models, where bright and dark solitons, periodic waves, and Jacobi elliptic-function solutions have been reported ~\cite{hasan2026exact, rabie2026exact, rateb2026analytical, rabie2026dual}}. Previous studies have shown that longitudinal engineering of the gain, GVD and nonlinearity in optical fibres significantly enhances the stability of the propagation of solitons and periodic waves. In particular, Ref. \citep{mani2016dynamics} has demonstrated the multi-soliton propagation in a tapered erbium-doped fibre under the action of a periodically distributed amplification. The foundational work\citep{1072858} has revealed that the solitons propagate with a minimal distortion in long fibre-transmission lines with periodically compensated losses. Work \citep{zheng2012propagation} addressed the behavior of chirped solitons in periodically
amplified variable-coefficient systems, and Ref.~ \citep{liao1999high} showed that the high-bit-rate soliton transmission can be maintained by the combination of the distributed gain and DM. In addition, Ref. \citep{mahalingam2009propagation} reported the controlled propagation of dispersion—and-nonlinearity—managed solitons in inhomogeneous erbium-doped fibre systems. Recent progress, including the development of the nonlinear Fourier transform, extends the soliton theory to encode transmitted data directly
into the nonlinear spectrum of the transmitted optical signals\citep{turitsyn2017nonlinear}. These studies again corroborate that the nonlinearity offers the foundation for the soliton-based transmission strategies, which are able to improve the performance of modern optical communications. Related studies explored the option of designing novel photonic-crystal fibres, with the nonlinearity strength varying exponentially along the propagation distance. They theoretically elaborated the generation and propagation of an array of optical pulses with constant gain parameters, using the Hirota bilinear-presentation technique \citep{atuba2016generation, manimegalai2016tapering}. Similarly, effects of the constant nonlinear and GVD terms in distributed-gain fibre amplification systems were investigated in Ref. \citep{atuba2021pulse}.

In this study, we consider the transmission of temporal optical signals in periodically amplified systems with fixed GVD and exponentially increasing or decreasing Kerr nonlinearity. {Unlike the general similarity-transformation treatments of distributed-coefficient NLSEs
elaborated in Refs.~\cite{kruglov2003exact, kruglov2005exact}, and the analysis of chirped solitary waves in periodically amplified variable-coefficient systems in Ref.~\cite{zheng2012propagation}, the present work considers fixed GVD combined with the exponentially increasing or decreasing Kerr nonlinearity, with the associated gain profile prescribed by the integrability condition. Within this gain–nonlinearity-engineered configuration, the cn-, dn-, and sn-type Jacobi elliptic periodic-wave families and their bright- and dark-soliton limits are obtained in the framework of a unified framework, and their response to finite perturbations and deviations from exact integrability is examined.}

\section{Periodic-Wave Solutions}
\noindent To establish the foundation for this study, we begin with the general form of the inhomogeneous NLSE, which governs the propagation of the optical envelope field $A$ in the fibre with spatially varying linear and nonlinear parameters. The full NLSE includes variable coefficients of the GVD, $p(z)$, nonlinearity, $q(z)$, and gain or loss, $g(z)$, along the propagation distance $z$, as shown in Eq.~(\ref{CGL}),

\begin{equation}
\label{CGL}
i\frac{\partial{A}}{\partial{z}}+p(z)\frac{\partial^2{A}}{\partial{t}^2}+q(z)|A|^{2}A+ig(z){A}=0.
\end{equation}

\noindent where $t$ is the usual reduced-time coordinate \cite{kruglov2003exact}. We proceed by simplifying Eq.~(\ref{CGL}) to a form applicable to the specific scenario considered here, in which GVD coefficient $p$ is kept constant, while the nonlinearity coefficient varies as a function of $z$:

\begin{equation}
\label{CGL2}
i\frac{\partial{A}}{\partial{z}}+p\frac{\partial^2{A}}{\partial{t}^2}+q(z)|A|^{2}A+ig(z){A}=0.
\end{equation}

\noindent The analysis reported in Ref. \cite{kruglov2003exact, kruglov2005exact} demonstrates that Eq.~(\ref{CGL}) admits exact analytical solutions, the key insight being to recognise that an appropriate similarity transformation can be applied to convert the original inhomogeneous equation into the standard integrable NLSE with constant coefficients, making it possible to use the great variety of its exact solution \cite{kengne2021spatiotemporal}. Earlier studies had also demonstrated that the integrable version of Eq.~(\ref{CGL}) (which is singled out by condition (\ref{condition}), see below) is directly solvable by means of the inverse-scattering technique \cite{serkin2000novel, serkin2002exactly}, confirming the existence of stable soliton waveforms even under the action of the varying parameters.

\noindent In the integrable case, exact wavetrain and pulse solutions to Eq.~(\ref{CGL}) are looked for as

\begin{equation}
A(z,t)=U(z,t)\exp[i \phi(z,t)],
\label{self}
\end{equation}

\noindent where the real amplitude is  
\begin{equation}
U(z,t)=\frac{1}{\sqrt{1+c_{0}D(z)}}M\left[\frac{t-t_c}{1+c_{0}D(z)}\right]\exp\left(G(z)\right).
\label{realM}
\end{equation}
and the quadratic phase is
\begin{equation}
\phi(z,t)=a(z)+c(z)(t-t_c)^2
\label{temppo}
\end{equation}

\noindent This phase structure implies that the pulse may carry a linear chirp. In this $\text{ans\"{a}tz}$, parameters $t_c$, $D(z)$ and $G(z)$ represent, respectively, the temporal position of the pulse's centre, the accumulated GVD, and the cumulative gain along the propagation distance, $viz.$,


\begin{equation}
D(z)=4\int_0^zp(z')dz'
\label{dz}
\end{equation}

\begin{equation}
G(z)=-\int_0^zg(z')dz'.
\label{TT}
\end{equation}

\noindent The variables $a(z)$, and $c(z)$ in Eqs.~(\ref{23}) and ~(\ref{gammm}) represent the evolution of the overall phase and linear chirp in the pulse:

\begin{equation}
a(z)=a_0+\lambda\int_0^z\frac{p(z')dz'}{[1+c_0D(z')]^2}
\label{23}
\end{equation}

\begin{equation}
c(z)=\frac{c_0}{1+c_0D(z)}
\label{gammm}
\end{equation}

\noindent In these expressions, $c_{0}$ represents the initial chirp parameter imposed at the fibre input, $a_{0}$ and $\lambda$ being constants. Further, function $M$ in Eq.~(\ref{realM}), which characterises the pulse envelope, is to be determined based on the specific variable profile of the nonlinearity coefficient $q(z)$ in Eq.(\ref{CGL}). In this work, we aim to construct cnoidal waveforms expressed in terms of Jacobi elliptic functions, which enable to obtain a broad family of nonlinear, quasi-periodic pulse trains.

 The validity of the self-similar transformation applied to the inhomogeneous NLSE is stipulated by the integrability condition, which ensures that the evolving pulse maintains its characteristic shape, while allowing the amplitude and phase to scale in a controlled manner:
 
\begin{equation}
g(z)=\frac{2c_0p(z)}{[1+c_0D(z)]}-\frac{1}{2}\left[\frac{q(z)}{p(z)}\right]\frac{d}{dz}\left(\frac{p(z)}{q(z)}\right)
\label{condition}
\end{equation}
 Thus, Eq.~(\ref{condition}) is the key requirement linking the physical design of the fibre system to the theoretical framework making its dynamics integrable \citep{kruglov2003exact, kruglov2005exact}. 
 
 In what follows below, We consider two cases defined by the following choices of the nonlinearity and gain modulation profiles in Eq.~(\ref{CGL}), while GVD coefficient, $p$ is fixed to be constant:


Case 1

\begin{equation}
q(z)=q_0\exp(-2\gamma z),\  \ g(z)=\frac{2c_0p}{1+4c_0pz} -\gamma ,
\label{condition1}
\end{equation}

 and Case 2,
 
 \begin{equation}
q(z)=q_0\exp(2\gamma z),\ \  g(z)=\frac{2c_0p}{1+4c_0pz} + \gamma ,
\label{condition2}
\end{equation}

\noindent where $\gamma$ is a real parameter. Both cases defined by Eqs~(\ref{condition1}) and (\ref{condition2}) satisfy the integrability condition (\ref{condition}). For these cases, the phase and chirp parameters $a(z)$ and $c(z)$, given by Eqs.~(\ref{23}) and~(\ref{gammm}), take the form of
\begin{equation}
a(z)=a_0+ \lambda\left[\frac{pz}{1+4c_0 pz} \right]
\label{phaf}
\end{equation}
\begin{equation}
c(z)=\frac{c_0}{1+4c_0pz}
\label{phafc1}
\end{equation}


\noindent { Introducing the similarity variable
\begin{equation}
\eta=\frac{t-t_c}{1+4c_0pz},
\end{equation}
the substitution of Eqs.~(3)--(9) into Eq.~(2), subject to the
integrability condition (10), reduces the governing equation to the
following nonlinear ordinary differential equation for the waveform
function $M(\eta)$:
\begin{equation}
\frac{d^2M}{d\eta^2}
+\frac{q_0}{p_0}M^3
-\lambda M=0.
\label{eq:reduced_M}
\end{equation}
The periodic-wave solutions of Eq.~(\ref{eq:reduced_M}) can be expressed
in terms of Jacobi elliptic functions. Accordingly, the corresponding
physical amplitudes $U(z,t)$ are given below.
}


\noindent For $pq>0$, we introduce two periodic solutions expressed in terms of the Jacobi elliptic functions with modulus $k$, $viz.$ (they are often called cnoidal waves because of the affinity to the basic elliptic function cn),
\begin{itemize}
    \item when $\lambda=(2-k^2)/t_0^2$
    \begin{equation}
U(z,t)=kA_c(z)\cn\left(\frac{t-t_c}{W(z)},k  \right),
\label{ampli12}
\end{equation}
    \item when $\lambda=(2k^2-1)/t_0^2$
    \begin{equation}
U(z,t)=A_c(z)\dn\left(\frac{t-t_c}{W(z)} ,k  \right).
\label{ampli13}
\end{equation}
    \end{itemize}

\noindent For $pq<0$ and $\lambda=-(1+k^2)/t_0^2$, the exact solution is

\begin{equation}
U(z,t)=kA_c(z)\sn\left(\frac{t-t_c}{W(z)} ,k \right).
\label{ampli14}
\end{equation}

\noindent Here $k$ is the elliptic modulus, the peak amplitude is $A_c(z)$, $t_0$ is the initial width. 

  
\begin{equation}
W(z)=t_0(1+4c_0pz).  
\label{wiw1}
\end{equation}

\noindent and the amplitudes for the two cases, which are defined above, are:

Case 1,
 \begin{equation}
A_c(z)= e^{\gamma z}\sqrt{\frac{2p}{q_0}}\frac{1}{t_0(1+4c_0pz)},
\label{ampli22a}
\end{equation}

Case 2,
 \begin{equation}
A_c(z)=e^{-\gamma z}\sqrt{\frac{2p}{q_0}}\frac{1}{t_0(1+4c_0pz)},
\label{ampli22b}
\end{equation}

\noindent Finally, collecting the above expressions, we arrive at the full expressions for the cnoidal-wavetrain solutions based on the three elliptic functions, cn, dn, and sn, in the two cases defined above (the respective solutions are denoted as $A_1$ and $A_2$).

\begin{itemize}

\item Case 1: The exponentially decreasing nonlinearity,

\begin{equation}
\begin{split}
A_1(z,t)=ke^{\gamma z}\sqrt{\frac{2p}{q_0}}\frac{1}{t_0(1+4c_0pz)}&\cn\left(\frac{t-t_c}{t_0(1+4c_0pz)}  \right)   \times \\
&\exp\left\{ia_0+ i\left(\frac{2k^2-1}{t_0^2}\right)\left[\frac{pz}{1+4c_0 pz}   \right]+ \frac{i c_0 (t-t_c)^2}{1+4c_0pz}\right\}
\label{ampli1112}
\end{split}
\end{equation}
\begin{equation}
\begin{split}
A_1(z,t)=e^{\gamma z}\sqrt{\frac{2p}{q_0}}\frac{1}{t_0(1+4c_0pz)}&\dn\left(\frac{t-t_c}{t_0(1+4c_0pz)}  \right)   \times \\
&\exp\left\{ia_0+ i\left(\frac{2-k^2}{t_0^2}\right)\left[\frac{pz}{1+4c_0 pz} \right]+ \frac{i c_0 (t-t_c)^2}{1+4c_0pz}\right\}
\label{ampli11122}
\end{split}
\end{equation}

{
\begin{equation}
\begin{split}
A_1(z,t)=ke^{\gamma z}\sqrt{\left|\frac{2p}{q_0}\right|}\frac{1}{t_0(1+4c_0pz)}&\sn\left(\frac{t-t_c}{t_0(1+4c_0pz)}  \right)   \times \\
&\exp\left\{ia_0 - i\left(\frac{1+k^2}{t_0^2}\right)\left[\frac{pz}{1+4c_0 pz}  \right]+ \frac{i c_0 (t-t_c)^2}{1+4c_0pz}\right\}
\label{ampli11123a}
\end{split}
\end{equation}
}

\end{itemize}

\begin{itemize}
\item Case 2: The exponentially increasing nonlinearity,

\begin{equation}
A_2(z,t)=e^{-2\gamma z} A_1(z,t) .
\label{A2}
\end{equation}
\end{itemize}


\noindent Solution~(\ref{A2}) for Case 2 actually exhibits three different eigenmodes that correspond to the solutions of the cn, dn and sn types in Case 1, which are presented by Eqs.~~(\ref{ampli1112}),~(\ref{ampli11122}) and~(\ref{ampli11123a}). In the long-wave limit, Case 1 amounts to the hyperbolic functions sech and tanh, that correspond, respectively, to the bright and dark solitary-wave solution of Eq.~(\ref{CGL}):

\begin{equation}
\begin{split}
A_1(z,t)=e^{\gamma z}\sqrt{\frac{2p}{q_0}}\frac{1}{t_0(1+4c_0pz)}&\sech\left(\frac{t-t_c}{t_0(1+4c_0pz)}  \right)   \times \\
&\exp\left\{ia_0+ \left(\frac{i}{t_0^2}\right)\left[\frac{pz}{1+4c_0 pz}   \right]+ \frac{i c_0 (t-t_c)^2}{1+4c_0pz}\right\}
\label{sech}
\end{split}
\end{equation}

{
\begin{equation}
\begin{split}
A_1(z,t)=e^{\gamma z}\sqrt{\left|\frac{2p}{q_0}\right|}\frac{1}{t_0(1+4c_0pz)}&\tanh\left(\frac{t-t_c}{t_0(1+4c_0pz)}  \right)   \times \\
&\exp\left\{ia_0- \left(\frac{2i}{t_0^2}\right)\left[\frac{pz}{1+4c_0 pz} \right]+ \frac{i c_0 (t-t_c)^2}{1+4c_0pz}\right\}
\label{tanh}
\end{split}
\end{equation}
}

\begin{figure}[htbp]
\myfigure
\centering{\includegraphics[width=12 cm]{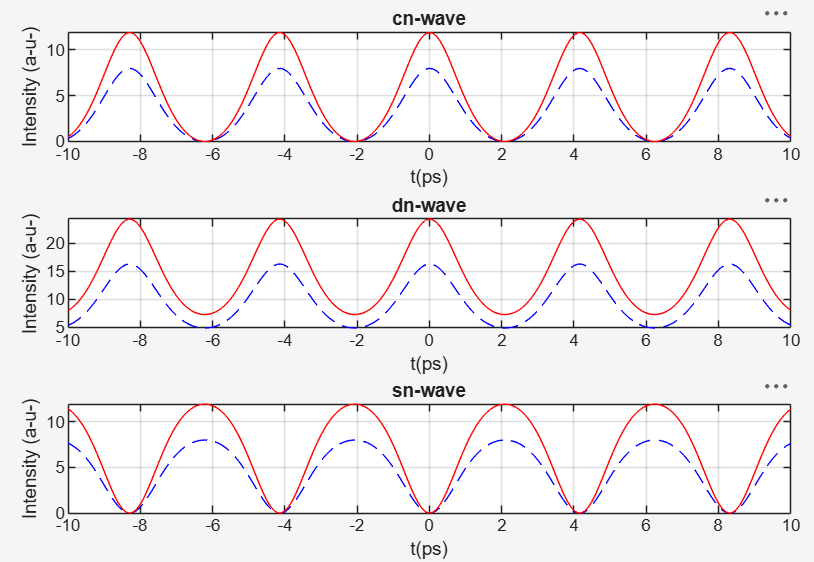}}
\caption{Intensity profiles of the initial (dashed lines) and amplified output (solid lines) of the cnoidal waves of the cn, dn, and sn types for Case 1, corresponding to the exponentially decreasing-nonlinearity {(nonlinearity-attenuation)} regime, as per Eqs.~(\ref{ampli1112}), (\ref{ampli11122}) and (\ref{ampli11123a}). The physical parameters are $p$ = $-$0.0553713 ps$^{2}$/m, $q$ = 0.006794 W$^{-1}$m$^{-1}$, $t_0=1~ps$, $tc=0.01~ps$, $a_0=0.03$, $c_0=0.03$~Hz/ps, $L_{\mbox{\scriptsize{D}}}$=9.03~m, $\gamma =0.00398355$m$^{-1}$, $z=2L_{\mbox{\scriptsize{D}}}$ and $k = 0.7$.}
\label{case1}
\end{figure}

\begin{figure}[H]
\myfigure
\centering{\includegraphics[width=12 cm]{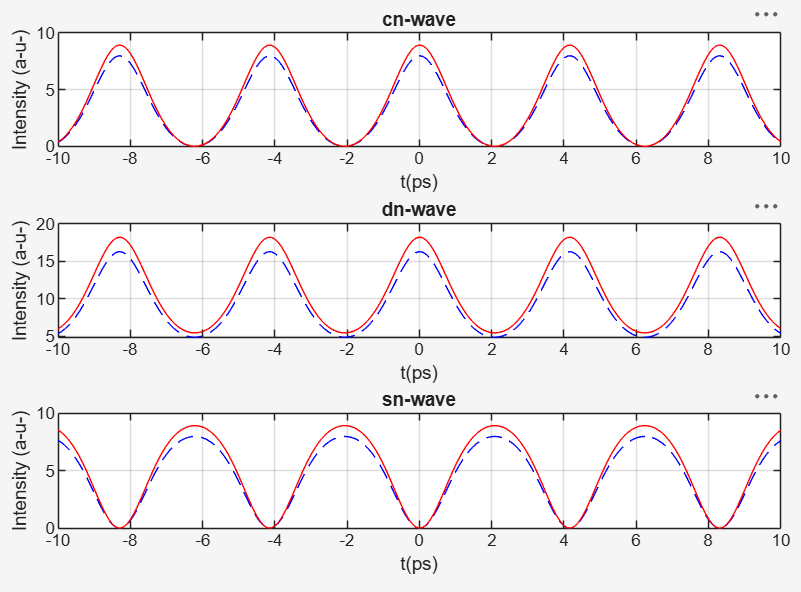}}
    \caption{Intensity profiles of the initial (dashed lines) and amplified output (solid lines) cnoidal waves of the cn, dn, and sn types for Case 2, corresponding to the exponentially increasing-nonlinearity {(nonlinearity-amplification)} regime, as per Eq.~(\ref{A2}). The physical parameters are $p$ = $-$0.0553713 ps$^{2}$/m, $q$ = 0.006794 W$^{-1}$m$^{-1}$, $t_0=1~ps$, $tc=0.01~ps$, $a_0=0.03$, $c_0=0.03$~Hz/ps, $L_{\mbox{\scriptsize{D}}}$=9.03~m, $\gamma =0.00398355$m$^{-1}$,  $z=2L_{\mbox{\scriptsize{D}}}$ and $k = 0.7$.}
\label{case2}
\end{figure}

\noindent {Although Eq~\ref{CGL2} may admit other Jacobi elliptic-function solutions, such as the nd-, sd-, and cd-type families \cite{zhaqilao2020nonlinear}, the present analysis is restricted to the  singularity-free cn-, dn-, and sn-type solutions because of their direct connection to physically relevant solitary-wave states. In the long-wave limit \(k\rightarrow1\), the cn- and dn-type solutions reduce to the bright-soliton \(\mathrm{sech}\) profile, whereas the sn-type solution reduces to the dark-soliton \(\tanh\) profile. These limit solutions are of particular relevance to the nonlinear fibre-optic systems considered in the present study.}

Figures \ref{case1} and \ref{case2} demonstrate that the pulse width remains essentially invariant for both exponentially increasing and decreasing nonlinearity profiles, despite the notable evolution in the pulse intensity over the considered length. This behaviour is explained by the fact that the evolution of the pulse is predominantly governed by the GVD term (See Eq.~(\ref{wiw1})); the GVD parameter being fixed, the pulse broadening or compression does not manifest itself. As a result, the wavetrain preserves both its temporal-envelope scale and phase chirp in the course of the propagation, indicating that the variation of the nonlinearity primarily affects the optical intensity and nonlinear phase accumulation, rather than the pulse duration. \noindent {Here, the constant GVD coefficient is defined as \(p=\beta_2/2\), where \(\beta_2\) is the second-order dispersion parameter. Accordingly, the characteristic dispersion length, \( L_D=\frac{t_0^2}{|\beta_2|} =\frac{t_0^2}{2|p|}\). For \(t_0=1~\mathrm{ps}\) and \(p=-0.0553713~\mathrm{ps^2/m}\), this gives \(L_D\simeq9.03~\mathrm{m}\). Hence, the maximum propagation distance \(z=2L_D\) considered in the figures correspond to \(18.06~\mathrm{m}\).}

As is well known \citep{agrawal2000nonlinear}, the nonlinearity coefficient in optical fibres is defined as

\begin{equation}
{q} = \frac{2\pi n_2}{\lambda A_\textrm{eff}}
\label{aeff}
\end{equation}

\noindent where $n_2$ is the nonlinear refractive index coefficient (m$^2$W$^{-1}$), $\lambda$ is the wavelength (measured in m), and A$_\textrm{eff}$ is the effective modal area  (measured in m$^2$). Equation~(\ref{aeff}) stresses that the engineered longitudinal variation of the fibre’s nonlinearity along the propagation distance, which is assumed in Eqs.~(\ref{condition1}) and (\ref{condition2}), may be implemented through the properly designed variation of the effective modal area. {For \(\gamma=0.00398355~\mathrm{m}^{-1}\), the assumed exponential profiles correspond to approximately a 15.5\% increase or a 13.5\% decrease ($\frac{q(2L_D)}{q_0}=e^{\pm 2\gamma(2L_D)}\approx e^{\pm 0.1445}$), respectively, in the nonlinear coefficient over the maximum propagation distance of \(18.06~\mathrm{m}\) (\(2L_D\)), representing relatively modest longitudinal variations. The physical magnitude of the exponential variation may be estimated from Eq.~\ref{aeff}. Assuming that \(n_2\) and the operating wavelength remain approximately constant, \(q(z)\propto A_{\rm eff}^{-1}(z)\); hence, for Case 2 (\(q(z)=q_0e^{2\gamma z}\)), the corresponding effective modal area varies as \(A_{\rm eff}(z)=A_{\rm eff}(0)e^{-2\gamma z}\), the effective modal area is reduced by approximately 13.4\% (\(A_{\rm eff}(z)/A_{\rm eff}(0) \approx 0.866\)). Under the approximate scaling \(A_{\rm eff}\propto r_{\rm eff}^2\), this corresponds to a reduction of about 6.9\% in the effective modal radius. Conversely, for the decreasing-nonlinearity profile, the effective modal area increases by approximately 15.5\%, corresponding to an approximately 7.5\% increase in the effective modal radius, such changes imply gradual taper-induced variations. This represents a relatively gradual longitudinal variation and is therefore physically plausible for a suitably designed tapered fiber. However, the relevant  mode confinement cannot be established from the value of $\gamma$ alone, as the confinement loss also depends on the specific fibre geometry, tapering ratio, operation wavelength, and tapering adiabaticity. Accordingly, the present value of $\gamma$ should be regarded as a physically reasonable parameter for the gradual tapering, rather than as one securing negligible confinement loss for arbitrarily tapered fibers. A quantitative assessment of the confinement loss would require a separate modal analysis of the specified fiber geometry, which is beyond the scope of the present analysis. }

For the exponentially increasing nonlinearity coefficient, the fibre may be tapered, so that the core diameter decreases along the fiber, leading to the enhanced mode confinement. When implemented in the combination with the periodically modulated optical amplification, this design sustains an elevated optical power along the fibre, enabling gradual enhancement/relaxation of the nonlinear effects, such as the spectral broadening and nonlinear phase modulation, while maintaining stable temporal profiles. This configuration also requires careful
management of the noise growth, and it must mitigate detrimental nonlinear impairments, including the stimulated Brillouin scattering.

Conversely, an exponentially decreasing nonlinearity profile can be produced by progressively increasing the core size or allowing the guided mode to expand into a larger area. Under the periodic amplification, this design concentrates strong nonlinear interaction near the fibre input, while preventing excessive nonlinear distortion or noise accumulation along the propagation path. This approach is suitable for scenarios requiring initial spectral shaping or phase manipulations with improved signal fidelity. Therefore, the ability to precisely control the distribution of the nonlinearity along the fiber provides an effective strategy for optimising the trade-off between the nonlinear efficiency, coherence preservation, and noise suppression. Examples of fibres exhibiting simultaneous exponential variation of both the GVD and nonlinearity, in the presence of a constant gain, have been demonstrated previously \citep{atuba2016generation}.

It is well established that optical pulses experience significant temporal reshaping when transmitted through a lossy nonlinear fibre, owing to the simultaneous power decay and nonlinear phase perturbation. Periodic amplification offers a solution for counteracting these detrimental effects, restoring the pulse’s energy at regular intervals and enabling stable long-distance transmission. The spacing between amplifiers, referred to as the gain length, is therefore a crucial design parameter. For the present study, the gain lengths, corresponding to the considered nonlinear configurations, can be estimated using Eqs.~(\ref{condition1}) and (\ref{condition2}). The concept of the guiding-centre soliton \citep{hasegawa1991guiding} provides the theoretical foundation for understanding the pulse evolution in such periodically amplified systems: the stable propagation can be maintained when the amplifier spacing remains short relative to the soliton period of the reshaped pulse emerging after each amplification event \citep{mollenauer1991long}. Longer amplifier separations are feasible when working with ultrashort, femtosecond-scale pulses, particularly if employing a gain medium capable of supporting substantial amplification factors \citep{malomed1994strong}. Indeed, peak intensities $\sim 1$~TW/cm$^2$  have been achieved for sub-60 fs pulses with total gains exceeding 1000, underscoring the viability of the high-power operation under the distributed-gain conditions. Distributed Raman amplification offers further advantages in this context, by providing continuous compensation of the fibre attenuation, while affording broad tunability in terms of the operating wavelength and flexibility in the pulse-duration control. Owing to the wavelength-dependent Raman gain spectrum, the amplification can be tailored to a wide range of signal wavelengths through appropriate selection of the pump wavelength \citep{headley2005raman}. This versatility makes the Raman-based distributed amplification an appropriate candidate for incorporation into the longitudinally-engineered nonlinear fibres, where the precise balance between the GVD, nonlinearity, and gain evolution is required.

\section{The Stability Analysis}
The crucially important feature of nonlinear  wavetrains is their stability against small perturbations \citep{choudhuri2016self}. Therefore, it is essential to assess the robustness of the generated chirped periodic wavetrain solutions with respect to finite initial perturbations. Such perturbations may arise from a variety of sources, including random noise, slight deviations in system parameters, or amplitude fluctuations \citep{choudhuri2012dark}.

\begin{figure}[H]
\myfigure
\centering
   \includegraphics[scale=0.5]{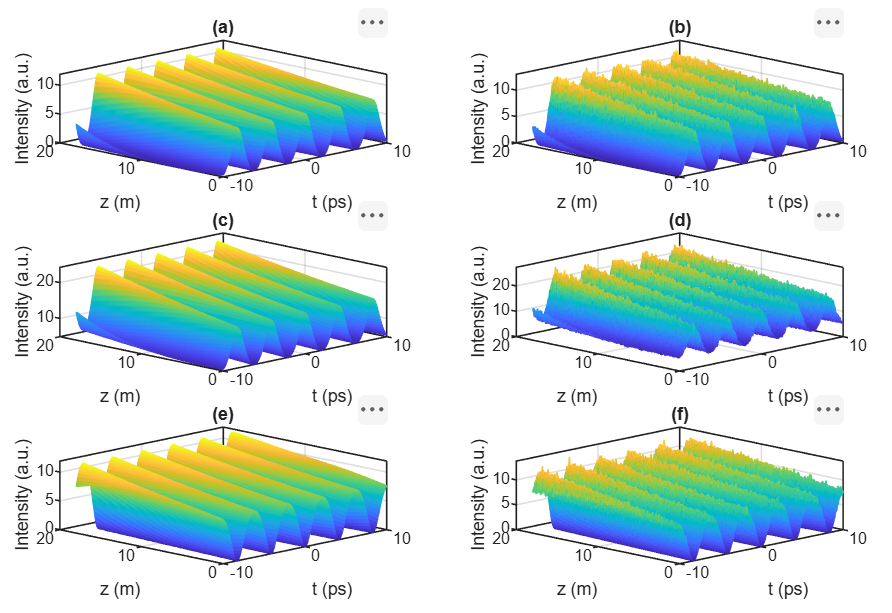}%
\caption{The evolution of the solutions of the (a) cn-wave, (c) dn-wave, and (e) sn-wave types under the action of the 10\% amplitude perturbation; the same for the (b) cn-wave, (d) sn-wave, and (f) dn-wave solutions under the action of the perturbation in the form of the 30~dB SNR additive white noise, in Case 1.  The physical parameters are  $p$ = $-$0.0553713 ps$^{2}$/m, $q$ = 0.006794 W$^{-1}$m$^{-1}$, $t_0=1~ps$, $tc=0.01~ps$, $a_0=0.03$, $c_0=0.03$~Hz/ps, $L_{\mbox{\scriptsize{D}}}$=9.03~m, $\gamma =0.00398355$m$^{-1}$,  $z=2L_{\mbox{\scriptsize{D}}}$ and $k = 0.7$. }
\label{case1SP}
\end{figure}

In the present study, the stability of the  cnoidal-wave solutions was examined by means of two complementary direct {computational} evaluation methods: considering amplitude perturbations and additive white noise. In the first case, the amplitude of the initial periodic wavetrain was perturbed at the 10\% level, to simulate potentially possible relatively large deviations in
the initial conditions. The resulting evolution for Case 1 is illustrated in Figs.~\ref{case1SP} (a), (c), and (e), which can
be directly compared to the unperturbed solutions shown in Fig.~\ref{case1}. The comparison demonstrates that the overall structure, periodicity, and phase characteristics of the wavetrain remain essentially unchanged, indicating strong resilience against the amplitude perturbations. 

\begin{figure}[H]
\myfigure
\centering
   \includegraphics[scale=0.5]{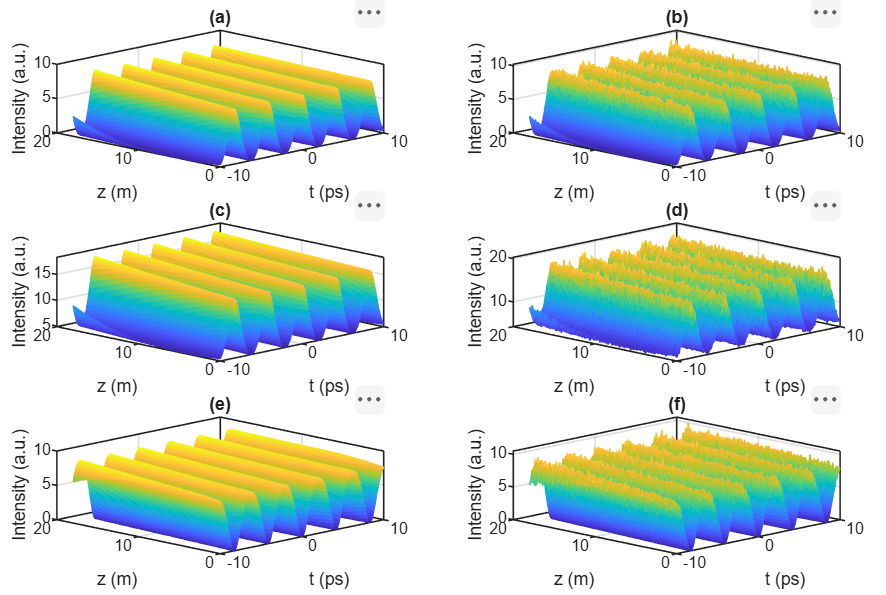}%
\caption{The evolution of the solutions of the (a) cn-wave, (c) dn-wave, and (e) sn-wave types under the action of the 10\% amplitude perturbation; the same for the (b) cn-wave, (d) sn-wave, and (f) dn-wave solutions under the action of the perturbation in the form of 30~dB SNR additive white noise in Case 2.  The physical parameters are  $p$ = $-$0.0553713 ps$^{2}$/m, $q$ = 0.006794 W$^{-1}$m$^{-1}$, $t_0=1~ps$, $tc=0.01~ps$, $a_0=0.03$, $c_0=0.03$~Hz/ps, $L_{\mbox{\scriptsize{D}}}$=9.03~m, $\gamma =0.00398355$m$^{-1}$,  $z=2L_{\mbox{\scriptsize{D}}}$ and $k = 0.7$. }
\label{case2SP}
\end{figure}

In the second case, the initial wave was superimposed with
additive white noise corresponding to the signal-to-noise ratio (SNR) of 30~dB. The evolution of this noisy input is presented in Figs.~\ref{case1SP} (b), (d), and (f). The results reveal that the additive noise does not significantly affect qualitative features or temporal profile of the periodic wavetrains. Similar observations were made for Case 2, as shown in Fig.~\ref{case2SP}, confirming that the wavetrains maintain their shape and coherence under the action of both types of the perturbations.

\begin{figure}[H]
\myfigure
\centering
   \includegraphics[scale=0.5]{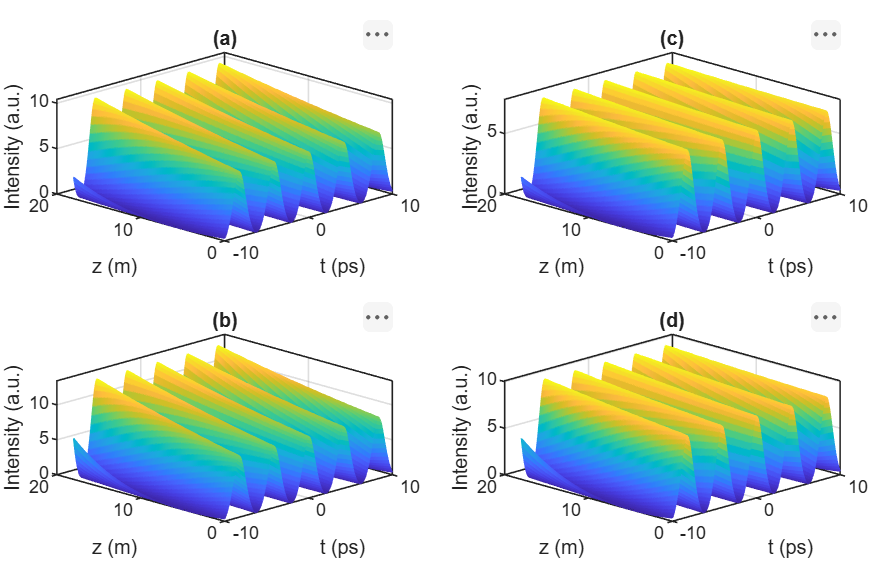}
\caption{The structural-stability analysis for the cn-wave solutions, tested by the simulations of their evolution in the case when the GVD parameter is $10\%$ different form the value necessary for the integrability (condition \ref{condition}). (a) Case 1 (-10\% GVD deviation); (b) Case 1 (+10\% GVD deviation); (c) Case 2 (-10\% GVD deviation); (d) Case 2 (+10\% GVD deviation).  The physical parameters are  $p$ = $-$0.0553713 ps$^{2}$/m, $q$ = 0.006794 W$^{-1}$m$^{-1}$, $t_0=1~ps$, $tc=0.01~ps$, $a_0=0.03$, $c_0=0.03$~Hz/ps, $L_{\mbox{\scriptsize{D}}}$=9.03~m, $\gamma = 0.00398355$m$^{-1}$,  and $z=2L_{\mbox{\scriptsize{D}}}$}
\label{StructuralPlot}
\end{figure}

The structural stability of the reported cn-wave solution is shown in Fig.~\ref{StructuralPlot}, It was tested by direct simulations against $\pm$ 10 $\%$  deviations of the GVD coefficient from the integrability condition in Eq.~(\ref{condition}), for cases 1 and 2 (which are defined above). These findings indicate that the wavetrain solutions, expressed in terms of the Jacobi elliptic functions of the cn, sn, and dn types, are completely stable, against the amplitude fluctuations, finite additive noise, and the deviation from the exact integrability. 

\begin{figure}[H]
\myfigure
\centering
   \includegraphics[scale=0.5]{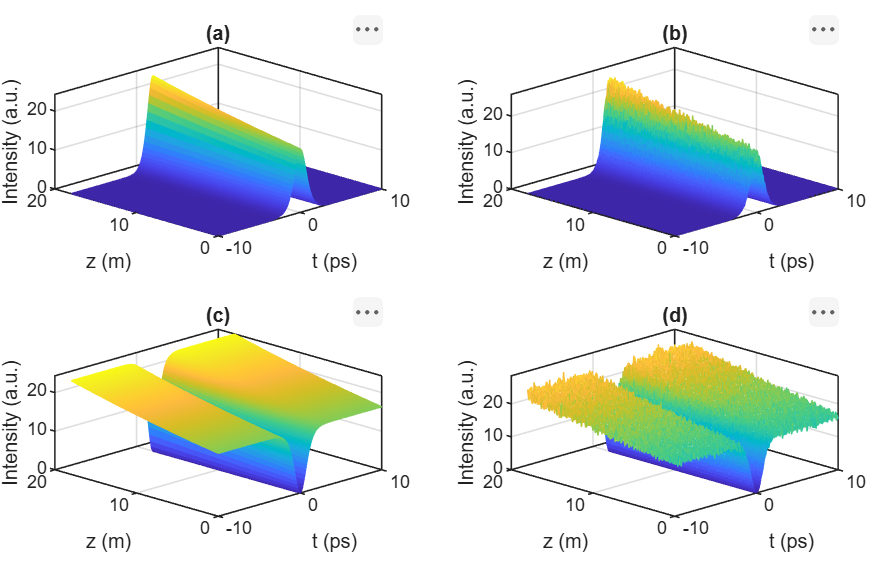}%
\caption{The evolution of the solitons of the (a) sech and (c) tanh types, under the action of the 10\% amplitude perturbation; the same for the (b) sech- and (d) tanh-type solitons under the action of the 30~dB SNR additive white noise in Case 1.  The physical parameters are  $p$ = $-$0.0553713 ps$^{2}$/m, $q$ = 0.006794 W$^{-1}$m$^{-1}$, $t_0=1~ps$, $tc=0.01~ps$, $a_0=0.03$, $c_0=0.03$~Hz/ps, $L_{\mbox{\scriptsize{D}}}$=9.03~m, $\gamma =0.00398355$m$^{-1}$,  and $z=2L_{\mbox{\scriptsize{D}}}$. }
\label{case1SS}
\end{figure}

\begin{figure}[H]
\myfigure
\centering
   \includegraphics[scale=0.5]{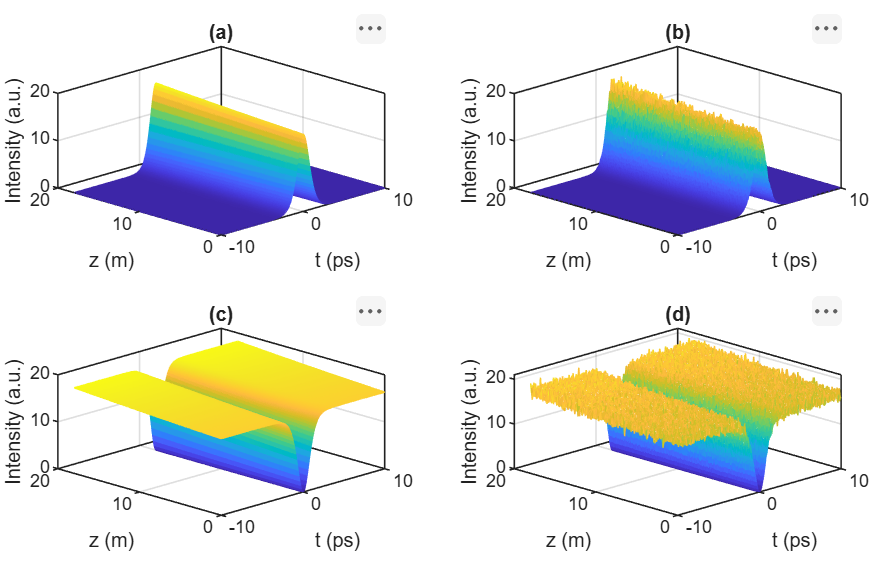}%
\caption{The evolution of the solitons of the (a) sech and (c) tanh types under the action of the 10\% amplitude perturbation; (b) the same for the sech- and (d) tanh-type solitons under the action of the perturbation in the form of the 30~dB SNR additive white noise in Case 2.  The physical parameters are  $p$ = $-$0.0553713 ps$^{2}$/m, $q$ = 0.006794 W$^{-1}$m$^{-1}$, $t_0=1~ps$, $tc=0.01~ps$, $a_0=0.03$, $c_0=0.03$~Hz/ps, $L_{\mbox{\scriptsize{D}}}$=9.03~m, $\gamma = 0.00398355$m$^{-1}$,  and $z=2L_{\mbox{\scriptsize{D}}}$}
\label{case2SS}
\end{figure}

Furthermore, the perturbed evolution of the dark and bright soliton solutions with the sech and tanh profiles (given by Eqs.~(\ref{sech}) and ~(\ref{tanh}), respectively), are displayed in Figs.~\ref{case1SS} and~\ref{case2SS}, further corroborating the overall stability of the fibre-optical modes obtained in this study. The stability against realistic perturbations underscores the practical relevance of these solutions for high-fidelity pulse propagation and control of nonlinear wavetrains in fibre systems.

\section{Conclusion}
In this work, we have systematically derived the families of exact wavetrain solutions for optical fibres with the distributed amplification, incorporating longitudinally varying nonlinearity and constant GVD (group-velocity dispersion). The modulation of these parameters along the propagation distance is chosen so that the underlying generalized NLSE is integrable, being gauge-equivalent to the classical NLSE with constant coefficients. The solutions are expressed in terms of the Jacobi elliptic functions of the sn, cn, and dn types (therefore, they are called cnoidal waves), providing a versatile framework for modeling the nonlinear-wave evolution in fibres with engineered parameter profiles. In the long-wave limit, these periodic solutions naturally reduce to the chirped solitons, highlighting the continuity between the periodic and solitary-wave regimes.

The analysis of the wavetrain dynamics reveals that their width scales remain constant, indicating that the temporal evolution is primarily governed by the constant GVD, while the longitudinally tailored nonlinearity drives the evolution of the optical intensity and nonlinear phase accumulation. This setup ensures the preservation of the wavetrains’ shape and their linear chirp in the course of the propagation, which may be of interest for applications requiring stable, high-fidelity wavetrains. The robustness of these solutions was verified by dint of extensive {computational evaluations}, considering finite initial perturbations of the amplitude or adding white noise. The structural stability against deviations from the exact- integrability condition was verified too. The results demonstrate that the qualitative features and periodic structure of the wavetrains are preserved under the action of the perturbations, confirming their resilience in the presence of imperfections and fluctuations in the input signal, and in the system’s scheme. These findings establish that the chirped periodic waves, as well as the solitons corresponding to their long-wave limit, can propagate stably in realistic fibre environments, making them promising candidates for applications to the supercontinuum generation, nonlinear pulse shaping, and high-power fibre-optic systems with distributed gain.

Overall, the present study provides an improved theoretical framework for understanding and utilising the longitudinally engineered nonlinearity in periodically amplified fibre systems. By combining the exact analytical solutions and stability analysis, it offers essential insights into the design and optimisation of nonlinear optical systems where controlled phase, amplitude, and spectral composition are required.
\bibliographystyle{unsrt}
\bibliography{references}
\end{document}